\documentclass[manuscript]{aastex}
\usepackage{emulateapj5}
\usepackage{psfig}

\slugcomment{submitted to ApJ Letters}

\shorttitle{The irradiated secondary of WZ Sge during the 2001 outburst}
\shortauthors{Steeghs et al.}

\begin{document}


\title{Emission from the secondary star in the old CV WZ Sge}


\author{D.Steeghs, T.Marsh, C.Knigge}
\affil{Department of Physics \& Astronomy, University of Southampton, Highfield, Southampton SO17 1BJ, UK}
\email{ds@astro.soton.ac.uk, trm@astro.soton.ac.uk, christian@astro.soton.ac.uk}
\author{P.F.L.Maxted}
\affil{Astrophysics Group, School of Chemistry \& Physics, Keele University, Staffordshire, ST5 5BG, UK}
\email{pflm@astro.keele.ac.uk}
\author{E.Kuulkers}
\affil{Space Research Organization Netherlands, Sorbonnelaan 2, 3584 CA Utrecht, NL}
\affil{Astronomical Institute, Utrecht University,
P.O. Box 80000, 3508 TA Utrecht, The Netherlands}
\email{erikk@purple.sron.nl}
\author{W.Skidmore}
\affil{School of Physics \& Astronomy, University of St.Andrews, North Haugh, St.Andrews, Fife KY16 9SS, UK}
\email{ws9@st-and.ac.uk}
\begin{abstract}

We present the  first  detection of the  mass  donor star in  the  
cataclysmic variable WZ  Sge.   Phase resolved  spectroscopy  reveals
narrow Balmer emission components  from the irradiated secondary  star
during the   2001  outburst.  Its  radial  velocity curve  indicates a
systemic  velocity   of $-72 \pm 3$  km/s  and   an  apparent velocity
amplitude of $K_{2_{app}}=493 \pm 10$ km/s.  
Doppler  tomography   reveals  a  highly   asymmetric accretion   disc
including a  significant bright  spot  contribution  20 days  into the
outburst. We estimate the primary radial velocity $K_1$ using a center
of symmetry method and find $K_{1_{app}}=37 \pm 5$ km/s. 
Accounting for the likely systematic  errors affecting both $K_1$  and
$K_2$  measurements, we conservatively  derive $508 <  K_2 < 585$ km/s
and $K_1 < 37$ km/s. This  implies a massive  white dwarf with $M_1 >  0.77
M_\sun$. A  non-degenerate mass donor,  implying  WZ  Sge has not  yet
evolved  through its minimum orbital  period, is  not ruled out by our observations. This would require an improved estimate of $K_1$.
Together with the measured phase   offset between bright spot  eclipse
and inferior  conjuction of  the  secondary star,  we can  bracket the
allowed mass ratio ($q=M_2/M_1$) to  lie between 0.040 and 0.073. This
provides a firm upper limit  to the mass  of  the secondary of $M_2  <
0.10 M_{\sun}$.

\end{abstract}


\section{Introduction}

%In Roche-lobe overflowing  accreting binaries, dynamically stable mass
%exchange occurs between   the    mass  donor and the     accretor.  

In cataclysmic variables  (CVs), a white dwarf  accretes from a  low
mass  secondary  star through   Roche-lobe overflow.  Angular momentum
loss drives the binary system to progressively shorter orbital periods
until the donor star becomes   degenerate. Further mass exchange  then
results  in  an increase of the   orbital period of  the system, which
leads to a predicted minimum period  of $\sim$70 minutes through which
all CVs should eventually evolve (e.g. Kolb \& Baraffe 1999).
With an orbital period  of only 82 minutes, WZ  Sge is one of the very
few candidates  among the hundreds of known  CVs that may have already
evolved past this   minimum  period (Patterson 1998).   Its  quiescent
magnitude  of $V\sim15.5$   and  distance  of   approximately  45 pc
(Thorstensen 2001, astrometric parallax, private communication), make
it one of the lowest luminosity CVs known. Despite the accretion light
being faint enough  for  the white  dwarf to  dominate  in quiescence,
infrared spectroscopy has so far not shown any signs of the mass donor
star (Littlefair et al. 2000).
Instead of the 3-5 magnitude  outbursts that longer period dwarf novae
tend to undergo every few weeks to  months, WZ Sge's outbursts have an
amplitude of 7-8 magnitudes  and recur  on  a timescale of roughly  33
years.    
These facts suggest  that WZ Sge  is a highly evolved system,
in which gigayears  of mass transfer  have  converted a main  sequence
secondary star to a degenerate brown  dwarf. However, the absence of any direct
signatures of the secondary  have prevented a robust  determination of
the system parameters of this unique system. 

On  July 23rd 2001,  amateur  observers  reported a  sudden and  rapid
brightening  of WZ   Sge (Ishioka et   al.  2001),  indicating another
outburst had started, around 23  years after the previous outburst  in
1978, and therefore around 10 years earlier than anticipated.
Here we present phase resolved spectroscopy of WZ Sge in the first weeks of
its 2001 outburst, which reveals  a clear signature of the  irradiated
mass donor star for the first time.

\section{Observations and reduction}

The phase resolved   spectra of WZ   Sge were obtained  with  the 2.5m Isaac Newton Telescope (INT) and the 4.2m
William Hershel Telescope  (WHT) on   the island  of  La  Palma.   The
intermediate dispersion spectrograph on the   INT in conjunction  with
the R1200B grating delivered  a wavelength coverage of 3800-4950\AA~at
0.48\AA/pixel using  an  EEV CCD detector. 
On  August the  6th, 1022  spectra were obtained  in total  on the INT
using  13 s  exposures  between 20:46 and  22:14  UT and 00:14 - 04:36
UT.
On the  WHT, the dual arm
ISIS  spectrograph was used, covering 4220-4975\AA~on  the blue arm at
0.22\AA/pixel  (EEV CCD) and 6380-6775\AA~at   0.4\AA/pixel on the red
arm (TEK CCD).  The slit width was adjusted in order to project to 2-3
pixels on  the CCD.  Since  no sufficiently bright comparison star was
available that could  be aligned along the slit,  the slit was  always
kept close to the parallactic angle.
 On August 13th, 541  red arm spectra using  10 s exposures and 233
blue arm  15 s exposures were acquired  between 20:50  and 23:03 UT.
Frames were   de-biased  using the  overscan areas  of   the CCDs.   A
normalised median of tungsten exposures was  then constructed for each
night  and used to carry out  flat  field correction.  Finally, the WZ
Sge spectra were optimally extracted  (Marsh, 1989).  Regular arc lamp
exposures allowed us  to  establish an  accurate wavelength scale  for
each   spectrum through  interpolation between    the two nearest  arc
spectra. The individual spectra were normalised to the continuum level using a spline fit to selected continuum regions.

\section{Data analysis}
\begin{figure}
\psfig{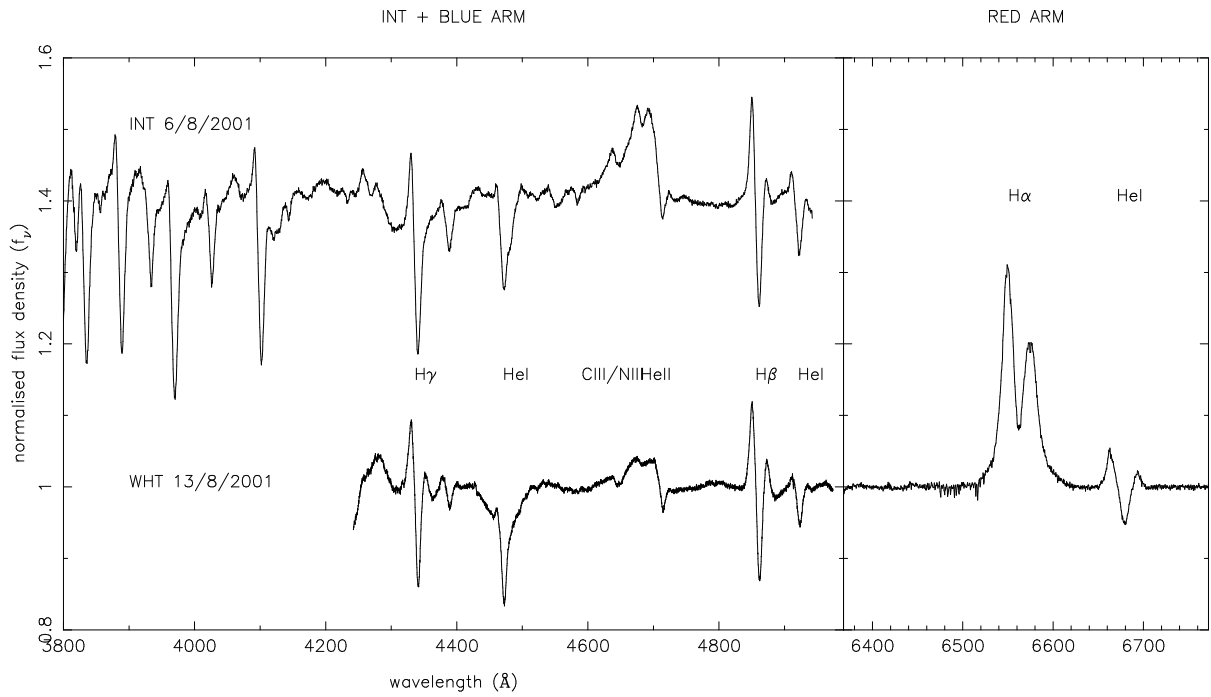}
\caption{The average spectrum of WZ Sge on August the 6th (top) and August the 13th. Spectra are normalised to the continuum and the blue INT spectrum of August the 6th is displaced upwards by 0.4. Several prominent lines are labelled.  \label{average}}
\end{figure}

 In  Figure \ref{average} we present the average normalised 
spectrum of WZ Sge  on  August 6th and 13th.   Both  observations
occured during  the initial slow  decline  phase of  the outburst with
approximate V   magnitudes of 10.0  on   the  6th (13  days  into  the
outburst)  and 10.5 on the 13th.  For comparison, $V\sim8$ at outburst
maximum     on   July      24th    (we    refer      to      the
AAVSO\footnote{http://www.aavso.org}                               and
VSNET\footnote{http://www.kusastro.kyoto-u.ac.jp/vsnet/}  webpages for
extensive visual magnitude estimates throughout the outburst).

The spectrum of WZ Sge in outburst  is dominated by complex Balmer and
Helium line profiles. 
Except for the HeII/Bowen blend  emission complex, all lines show very
deep and phase dependent absorption components on top of double peaked
emission. In  most cases the absorption goes  well below the continuum
level, except in H$\alpha$. 
%The HeII emission  line equivalent width is consistently
%dropping  on the linear  decline phase, accompanied by less pronounced
%double peaks in its line profile.
%
We use the  orbital ephemeris of Patterson  et al.   (1998), hereafter
P98,  to  calculate the  orbital phases throughout   this Letter.  The
ephemeris  is a small improvement  to the often-used Robinson, Nather
and  Patterson  (1978;   hereafter  RNP)  ephemeris   amounting  to  a
difference of $\sim$0.001 cycles at the  time of the current outburst.
The ephemeris zeropoint is  based on the sharp  eclipses of the bright
spot and does  not  correspond  to the  inferior conjunction  of   the
secondary  star.  Spruit \& Rutten  (1998),  hereafter SR, for example
derive a phase offset of $-0.041  \pm 0.003$ between the RNP ephemeris
and the mid-point of the accretion disc eclipse.  We will address this
phase offset in Section 3.3.

\subsection{Balmer emission from the secondary}

\begin{figure}
\psfig{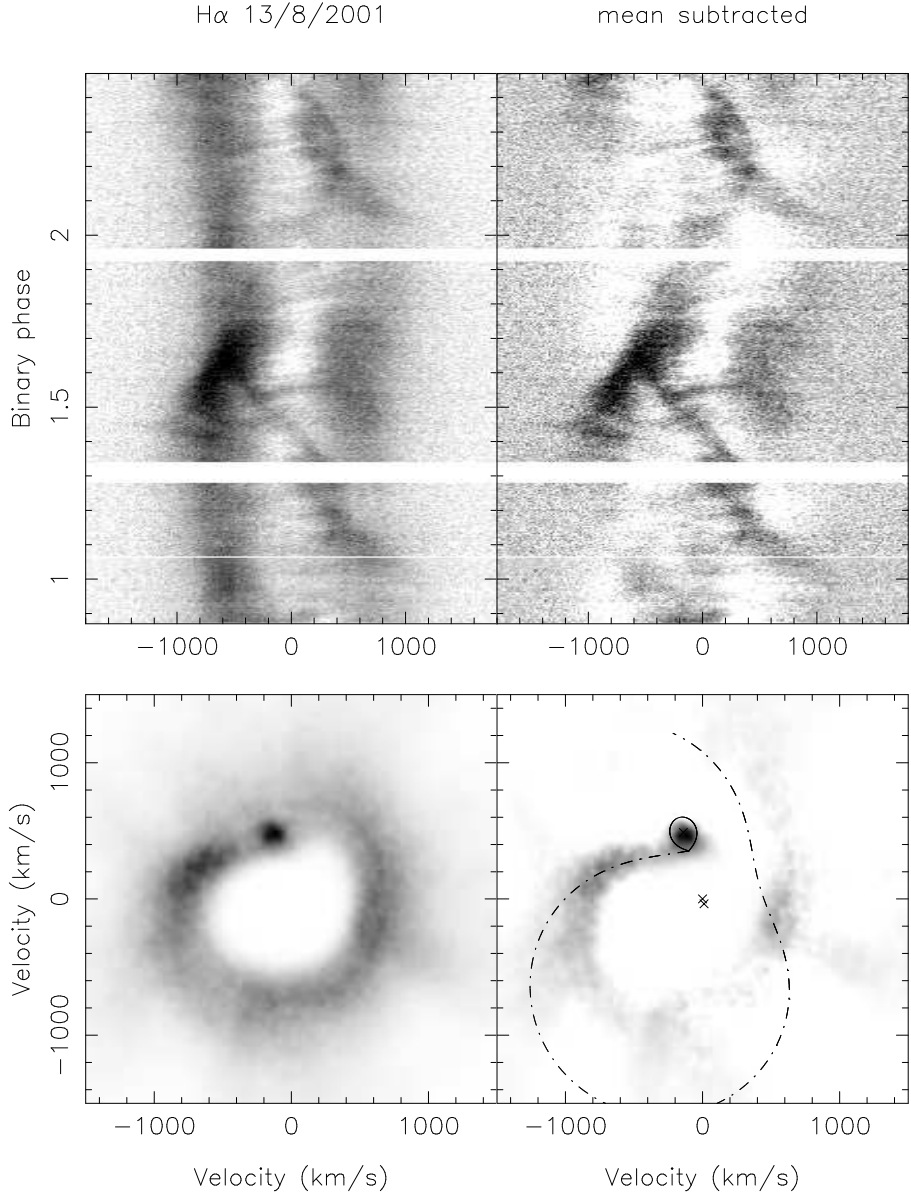}
\caption{Top left: the observed H$\alpha$ emission as a function of orbital phase on August 13th. Below, the corresponding Doppler map revealing the mass donor star in emission. Top right: the observed data after the mean spectrum was subtracted, highlighting the asymmetries in the accretion flow as well as the S-wave from the secondary. Bottom right is the asymmetric part of the H$\alpha$ tomogram, obtained through subtraction of the symmetric component centered on the expected location of the white dwarf. The predicted location of the Roche lobe and ballistic gas stream is plotted for $q=0.073$ ($K_2=508$ km/s, $K_1=37$ km/s) and $\Delta\phi_{spot}=0.046$.  \label{trails}}
\end{figure}

The time dependent H$\alpha$ line  emission as observed on August 13th
is  displayed in Figure  \ref{trails}  as a  trailed spectrogram.  The
double-peaked  line profile is  highly phase dependent and asymmetric.
Apart from the two double peaks reflecting emission from the accretion
disc, a narrow emission component moves through the peaks and produces
a clear S-wave  on  the orbital  period.  On much  shorter timescales,
rapid transient features can be seen crossing the line profile, almost
always from blue to red. An example is the narrow emission
feature rapidly crossing in a few  minutes around phase 0.55.
During  quiescence, the line  profiles of WZ Sge  are dominated by the
double-peaked disc emission  as well as  a strong contribution from  the
bright spot (SR,  Skidmore et al.  2000).  However, the S-wave in  our
outburst data  is much  narrower and  traces a near  sinusoidal radial
velocity curve.
Such narrow, sinusoidal   S-waves are commonly observed  in  accreting
binaries and are generally attributed to line  emission from an irradiated
secondary star. Provided a  sufficient amount of ionising radiation is
received from  the compact object  and the inner disc regions, optical
line emission from the exposed parts of the secondary star is produced
(e.g.  Marsh \& Horne 1990, Harlaftis et al.  1996, Steeghs
\& Casares, 2001).  This interpretation is supported  by the phasing of 
the S-wave, which  corresponds closely to the  expected phasing of the
secondary  in WZ  Sge. In   addition,  its strength   is highly  phase
dependent,  reaching  maximum emission around    orbital phase 0.5, as
expected from an irradiated Roche lobe filling star.
Given the complexity  of the line profiles, we  decided to use Doppler
tomography  (Marsh \& Horne  1988) to isolate  and study the nature of
the emission line components. 

\subsection{The systemic velocity}

Doppler tomography requires  the  systemic velocity ($\gamma$)  to  be
supplied as input to the reconstruction algorithm.
Gilliland  et al.  (1986)  measured   $\gamma=-72 \pm 3$ km/s  using  radial
velocities derived from  the H$\alpha$  line  wings.  Skidmore et  al.
(2000) derive a  mean systemic velocity of  $-78 \pm 9$ km/s  based on
radial   velocity curves of several   emission  lines.  Both of  these
determinations  rely   on  the   assumption that  the   emission  line
velocities reflect the  motion of the   compact object.  On  the other
hand, SR used Doppler tomography techniques  to determine the systemic
velocity, by minimising  the residuals between observed and  predicted
data. They found $\gamma = -71\pm 3$ km/s.

We chose to measure the systemic velocity directly using the detected
S-wave from the secondary star, since its radial velocity is displaced
by the true systemic velocity of the binary system irrespective of the
properties of the accretion flow around the white dwarf.
To this end we calculated a  series of preliminary Doppler  maps from the observed
data using  a filtered back  projection method (Marsh  2001). For each
map,  a different systemic velocity was  assumed ranging from 0 to -110
km/s in 10 km/s steps.  The Doppler  maps then provide the strength of
all S-waves on the orbital  period in the data  with a given amplitude
and phase, which is maximised when the correct value for $\gamma$ is used.
We measured  the  strength of  the   secondary star emission  in  each
Doppler image, and found that a  well defined maximum was  achieved for $\gamma =
-74   \pm 3$ km/s  using the  observed  H$\alpha$  emission.  The same
analysis  applied  to the  H$\beta$   emission  also reveals   a clear
secondary star contribution which is maximised for $\gamma = -69 \pm 3$
km/s.
%
%The secondary star emission in the higher Balmer lines is too weak and
%is affected by the strong  absorption for them to provide  additional,
%reliable constraints to the gamma velocity.
%
We thus use a systemic
velocity  of $\gamma= -72$  km/s throughout this  Letter, based on the
mean of the H$\alpha$ and H$\beta$ values.
Our systemic velocity, based on the radial  velocity curve of the mass
donor star,  is in  close agreement with  the values  determined from the
disc emission lines. 

\subsection{The radial velocity of the secondary}

The final Doppler tomogram  illustrating the distribution of H$\alpha$
emission  on  August 13th is   displayed  in Figure \ref{trails}.  The
tomogram was  constructed from a regularised  fit to the observed line
profiles,  using  maximum  entropy regularisation (Marsh  2001).   The
secondary star emission maps to a sharp  spot with a FWHM of $\sim$130
km/s compared  to a resolution element   of 36 km/s.  Maximum emission
occurs at $V_x=-140  \pm 10$ km/s,   $V_y=470 \pm 10$ km/s as  derived
from a 2D Gaussian fit.
If the data was folded on the correct orbital ephemeris, emission from
the mass donor should appear on the positive $V_y$-axis, corresponding
to the radial  velocity ($K_2$) of the  mass donor star.  The emission
of the secondary thus  allows us to   calculate the phase of  inferior
conjunction relative to the photometric ephemeris of P98. If we assume
that the center of the H$\alpha$ emission corresponds to the center of
the  mass donor we can derive  a phase offset of  $-17 \pm 1 ^{\circ}$
($\Delta\phi_{spot}=-0.046\pm 0.003$ in  terms of orbital  phase) and an apparent  radial velocity
amplitude of $K_{2_{app}} = 493 \pm 10$ km/s.   Our value for the phase offset
appears slightly  larger  than that of  SR  based on  the disc eclipse
during quiescence, but is still within 2 sigma of their value.
For comparison, the  same analysis applied to the  Doppler maps of the
H$\beta$ and $H\gamma$ emission  leads to identical phase  offsets for
both lines  ($17 ^{\circ}$) and radial  velocities  of $478\pm10$ km/s
(H$\beta$) and $479\pm10$ km/s  (H$\gamma$) respectively.  There is no
evidence for any  secondary star emission  in the HeI6678, HeII4686 or
Bowen blend transitions, indicating that  the secondary star is exposed
to relatively soft ionising radiation.
The  quoted uncertainties  on   these  values  does not include    the
systematic errors    that affect  both   $K_2$  and  the  phase  offset
because of the  unknown  distribution of the line
emission across the Roche lobe (c.f. Steeghs \& Casares, 2001). If the
line emission is biased towards either the left or right hemisphere of
the  lobe, a corresponding  bias to the  derived phase offset would be
introduced.
%Similarly,  if  the emission is  biased
%towards the  front half of the  Roche lobe rather, the  derived radial
%velocity amplitude  is  smaller than the  true radial  velocity of the
%center of mass of the secondary.
%
Given that only the  front part of  the Roche lobe is  irradiated, and
that no intrinsic line emission from  the secondary is observed during
quiescence, the  apparent radial  velocity amplitude of the emission $K_{2_{app}}$ will be smaller than the true radial velocity $K_2$ of the secondary. 
The observed  line emission  must originate  somewhere between the  L1
point and the   terminator that separates  the irradiated  part of the
Roche lobe from its unirradiated   side.  The conservative  assumption
that all emission  originates   at  the terminator implies  that   the
smallest possible correction between $K_{2_{app}}$ and $K_2$ is around
3\%.  This was  derived from Roche  geometry  calculations  across the
allowed mass ratio range.  Thus our detection of H$\alpha$ emission at
493 km/s implies $K_2 > 508$ km/s.
On the other hand, the observed velocities cannot  be smaller than the
velocity of the L1 point, which leads to an upper limit of $K_2 < 585$
km/s. Here we have again allowed for a  wide range of mass  ratios consistent with  $M_1<  1.4M_{\sun}$.
%

%\subsection{Comparison with additional spectroscopy}
%The Doppler  maps of WZ Sge on  August the 13th are markedly different
%from  those in the   first few days  of  the outburst.  Orbit resolved
%spectroscopy on July the 28th, only 5  days into the outburst revealed
%an accretion disc dominated  by two spiral arms  (Steeghs et al., IAUC
%7675), and no sign of  any secondary star  emission in either H$\beta$,
%HeI or  HeII (H$\alpha$ was not observed).  
%
%The first hints of secondary star emission  appeared in phase resolved
%spectroscopy  obtained on August     1st in H$\alpha$   and  H$\beta$,
%producing  a weak emission  spot  in the Doppler  tomograms, a feature
%that also consistently appeared in Doppler  maps produced from spectra
%obtained between the 6th and 9th of August.  The phase offsets derived
%from these  maps range from   16 to 18  degrees, with  apparent radial
%velocities   between 476  and  497 km/s.    All  of these  values  are
%consistent with the August 13.
%
%A combination of largest strength  of the secondary emission component
%and  relatively  little distortion  by  underlying
%absorption makes the August 13th H$\alpha$ data the best suited to date 
% for determining the properties of the mass donor.
%

\subsection{The radial velocity of the primary}
 
Armed with a  good estimate for  the radial velocity amplitude  of the
secondary star,  the mass ratio $q=M_2/M_1=K_1/K_2$  of the binary can
be determined if the  radial velocity of the  primary ($K_1$) is  also
known.  Gilliland et al.  (1986) obtained  an estimate of $K_1 = 48\pm
6 $ km/s  from the radial velocities of  both H$\alpha$ line wings and
peaks.
However, the radial velocity curves show phase offsets with respect to
the   absolute  ephemeris which indicate    that these radial velocity
curves must be severely distorted.  This  is a common situation in CVs,
and may  not be surprising given  the strong bright spot emission that
is present in WZ Sge during quiescence.
Mason et al.   (2000) also measured  emission  line velocities using  a
wide range of spectral lines in the  optical and infrared regime. They
found  velocity  amplitudes between  46 and  121 km/s   and large phase
offsets depending on the   excitation potentials of  the  lines.  They
concluded  that a varying degree  of bright spot contamination distorts
the radial velocity curves of the emission lines.
%
%Disc flow asymmetries and  non-Keplerian  flows increase the  apparent
%radial velocity amplitude of  the disc emission, and thus overestimate
%the radial velocity of the primary star.  

SR   used a different   approach and attempted to  find  the center of
symmetry of the disc emission in the  Doppler map at a given velocity,
ignoring areas that are affected by the bright spot. They found that at
large velocities the center of  symmetry seems to converge on $K_{1_{app}} =
40 \pm 10$   km/s,   even though  the phase    offset  is still  considerable
($50^{\circ}$).
We  applied a similar method to  the  outburst H$\alpha$ tomogram, and
find a convergence to a center of symmetry at $K_{1_{app}} = 37 \pm 5$
km/s   at high  velocities  (1200-1500  km/s)  before  noise starts to
dominate.  The optimal center of symmetry, like  in the case of SR, is
offset from the expected position of the  white dwarf corresponding to
a phase shift of $60^{\circ}$.  If we  force the center of symmetry
to be phased  with the white  dwarf while minimising  the residuals at
areas not affected by the bright spot, we find $K_{1_{app}} = 40 \pm 5$ km/s.
Although the formal uncertainty of  this optimal center of symmetry is
only a few km/s, our methods may  be affected by systematic errors due
to the fact that we are relying on a complicated emission structure to
reflect  the motion of  the white dwarf.   As indicated by our Doppler
images,  the accretion flow  is  highly asymmetric, and a  significant
amount of  distortion may be expected.   We therefore consider 37 km/s
to be an upper   limit to the  true  radial velocity amplitude  of the
white dwarf.

The bottom-right  panel of Figure \ref{trails} plots
the  asymmetric  part of the   H$\alpha$ emission, after the symmetric
part  with respect to the  optimal  center of symmetry was subtracted.
Significant asymmetries  are   clearly present,  and  the  resemblance
between our  outburst map and the  quiescent $H\alpha$  Doppler map of SR is
both striking and surprising. It appears a substantial contribution to
the line  flux  originates from  the bright  spot region. It  has been
proposed in the past (Smak 1996 ; Hameury, Lasota \& Hure, 1997), that
heating of the  secondary during outburst may  lead to an  increase in
the mass transfer rate and thereby prolong the outburst duration.
We refrain from  speculating about the nature  of the disc asymmetries
until a more thorough comparison with other outburst tomography throughout the 2001 campaign can be made.

\subsection{The system parameters}

White dwarf mass estimates have led to a wide range of published white
dwarf masses in WZ  Sge ranging from  0.3$M_\sun$ to 1.2$M_\sun$.  Our
lower limit to the radial velocity of the secondary star ($K_2 > 508$ km/s)
leads to a mass function of;

\[ f(M_1) = \frac{P K_2^3}{2\pi G} = \frac{M_1 \sin{i}^3}{(1+q)^2} > 0.77 M_\sun \]
Thus the low  white dwarf mass values of,
for  example, Smak   (1993),  RNP and   Cheng  et al.   (1997), are  not
compatible with our  $K_2$ measurements, and a  more massive white dwarf is
required. 
%Given that  WZ Sge shows disc  and bright spot eclipses, but
%the absence of white dwarf eclipses  reliably fixes the inclination to
%$75
%\pm 2  ^{\circ}$ (Smak 1993, SR). The lower limit for
%the white dwarf mass then becomes $M_1 > 0.79 M_\sun$.
%
%On the other hand the fundamental upper limit  to the white dwarf mass
%is the 1.4$M_\sun$ Chandresekhar limit. 
With  $K_2 >  508$ km/s  and  $K_1 < 37  $  km/s we  have  $q < 0.073$
implying   $M_2  < 0.10 M_\sun$ since    $M_1  < 1.4M_\sun$.   Thus, a
non-degenerate secondary star is formally not yet ruled out.
However, the lack of any contribution of the mass donor to the J and K
bands (Littlefair et al.  2000) is difficult  to reconcile with a late
main sequence  mass donor around  $\sim 0.1  M_\sun$.  Even for highly
evolved main sequence stars, the predicted J and K band magnitudes are
significantly too bright (e.g. Leggett et al. 2001).
If, as expected   for a system  that  has evolved through  the  period
minimum, the secondary  star is in  fact a degenerate star with  $M_2<
0.076M_\sun$, $K_1$ must be less than $28$ km/s.
This clearly illustrates the need for an accurate determination of the
radial velocity of  the primary in  WZ Sge.  Given that  the accretion
flow is clearly asymmetric both during quiescence as well as outburst,
this may only be possible through  the use of photospheric white dwarf
line velocities.   Cheng et  al. (1997) did   not detect a  systematic velocity shift in their HST data of WZ Sge.

The  measured phase   offset  ($\Delta\phi_{spot}$)   between inferior
conjunction of the secondary and bright  spot eclipse provides another
constraint to the allowed mass  ratio range.  We calculated gas stream
trajectories in order to  determine the  predicted $\Delta\phi_{spot}$
as  a function  of mass   ratio and  disc  radius.   Allowing for  the
uncertainty in disc   radius measurements, we  can then  rule out mass
ratios smaller than $q<0.040$  since the predicted phase  offset would
be  larger      than      0.049  compared    to     our   value     of
$\Delta\phi_{spot}=0.046\pm.003$, and $\Delta\phi_{spot}=0.041\pm.003$
as derived by SR.

\section{Discussion}

We have detected Balmer  emission originating from the irradiated mass
donor in the CV WZ  Sge during the second and  third weeks of its 2001
outburst.  This is  the first time a  direct detection of the low mass
secondary in  WZ Sge has been made. 
The Doppler maps of WZ Sge on  August 13th are markedly different from
those   in  the  first  few days    of the  outburst.   Orbit resolved
spectroscopy on July 28th, only  5 days into  the outburst revealed an
accretion disc dominated by two spiral arms (Steeghs et al. 2001), and
no sign of any secondary star emission in either H$\beta$, HeI or HeII
(H$\alpha$ was not observed).
By August  13th, not only is the  secondary star present  in emission,
the accretion flow also  has made a major  transition.
The disc emission is dominated by  a strong extended bright spot, very
similar to  its quiescent structure even though  the system is still 5
magnitudes brighter than its quiescent level.
The implications of this in terms  of varying mass transfer and bright
spot contribution throughout the 2001   outburst will be pursued in  a
future paper.

A reliable determination of  the component masses in  WZ Sge awaits an
accurate  determination  of the  radial velocity  of  the white dwarf. 
Accounting for the   possible systematic errors affecting both  $K_1$
and $K_2$  measurements, we conservatively  derive  $508 < K_2  < 585$
km/s, $K_1 <  37$ km/s  and $0.040  < q <   0.073$.  In terms  of component
masses, this corresponds  to $0.77 <  M_1  < 1.4M_\sun$,  while $M_2 <
0.10 M_\sun$.

\acknowledgments

DS is  supported by a PPARC Fellowship. 
The  William Herschel and Isaac  Newton telescopes  are operated on the
island  of  La   Palma  by  the  Isaac Newton   Group  in the  Spanish
Observatorio  del   Roque  de  los   Muchachos  of the    Instituto de
Astrofisica de Canarias.
Many thanks to the Isaac Newton Group  staff for obtaining part of the
data through the ING service programme. 
We would also like to thank the many collegues and amateur observers who are  contributing to the WZ Sge campaign during its 2001 outburst.

%\newpage

\clearpage

% FIGURE 1

% FIGURE 2

%\clearpage


\end{document}